\documentclass[12pt]{article}
\usepackage{amssymb}
\usepackage{graphics}
\usepackage{epsfig}
\usepackage{a4wide}

\begin{document}
\newcommand{\eettz}{$e^+e^- \to t \bar t Z^0$ }
\newcommand{\eettzg}{$e^+e^- \to t \bar t Z^0g$ }
\newcommand{\eettzga}{$e^+e^- \to t \bar t Z^0\gamma$ }

\title{ One-loop QCD and electroweak
corrections to $t\bar{t}Z^0$ production at an $e^+e^-$ linear
collider \footnote{Supported by National Natural Science Foundation
of China.}}
\author{ Dai Lei, Ma Wen-Gan, Zhang Ren-You, Guo Lei, and Wang Shao-Ming \\
{\small Department of Modern Physics, University of Science and Technology}\\
{\small of China (USTC), Hefei, Anhui 230026, P.R.China}  }

\date{}
\maketitle \vskip 15mm
\begin{abstract}
We study the impact of the full ${\cal O}(\alpha_{s})$ QCD and
${\cal O}(\alpha_{ew})$ electroweak (EW) radiative corrections to
the $e^+e^- \to t \bar t Z^0$ process in the standard model (SM),
and investigate the dependence of the lowest-order(LO), one-loop QCD
and EW corrected cross sections on colliding energy $\sqrt{s}$. The
LO, QCD and EW corrected spectrums of the invariant mass of $t\bar
t$-pair and the distributions of the transverse momenta of final
top-quark and $Z^0$-boson are presented. The numerical results show
that the one-loop QCD correction enhances the LO cross section, but
the EW one-loop correction generally suppresses the LO cross section
with our chosen parameters. In the case of $m_H=120~GeV$, the QCD
relative corrections can reach $43.16\%$ when $\sqrt{s}=500~GeV$,
while the EW relative corrections have the values of $-8.63\%$,
$-4.07\%$ and $-5.42\%$, when $\sqrt{s} = 500~GeV$, $800~GeV$,
$1.2~TeV$, respectively.

\end{abstract}


\vfill \eject

\baselineskip=0.32in

\renewcommand{\theequation}{\arabic{section}.\arabic{equation}}
\renewcommand{\thesection}{\Roman{section}.}
\newcommand{\nb}{\nonumber}

\newcommand{\Dir}{\kern -6.4pt\Big{/}}
\newcommand{\Dirin}{\kern -10.4pt\Big{/}\kern 4.4pt}
\newcommand{\DDir}{\kern -7.6pt\Big{/}}
\newcommand{\DGir}{\kern -6.0pt\Big{/}}

\makeatletter      
\@addtoreset{equation}{section}
\makeatother       

\section{Introduction}
\par
The standard model(SM)\cite{s1,s2} of elementary particle physics
has provided a remarkably successful description of almost all
available experimental data involving strong and electroweak (EW)
interactions. While on the contrary, the Higgs-boson in the SM
has not been discovered experimentally yet. That implies that the mechanism
of the electroweak symmetry breaking (EWSB) giving masses to
vector gauge bosons and fermions, remains a mystery. Moreover,
the SM suffers from a few conceptional difficulties, such as hierarchy
problem. This has promoted intense researches in building extension
models in order to solve the hierarchy problem, and leaded to a rich
phenomenology at present and future colliders.

\par
The top-quark was discovered by the CDF and D0 collaborations at
Fermilab Tevatron in 1995\cite{cdftop,d0top}. It opens up a new
research field of top physics, and confirms again the
three-generation structure of the SM. But until now our knowledge
about top-quark's properties has been still limited \cite{Chak}.
In particular, the couplings of the top-quark to the EW gauge
bosons have not yet been directly measured in high precision.
Since the top-quark is the heaviest particle discovered up to
now\cite{tew,hepdata}, it indicates that among all the observed
interactions of elementary particles, the top-quark mass term
breaks the EW gauge symmetry maximally, and the detailed physics
of the top-quark may be significantly different from the
predictions provided by the SM. In other words, the top-quark
probably plays a special role in EWSB. The studies of top physics
could illuminate the mechanism which breaks EW symmetry. Thus we
may guess the new physics connected with EWSB could be firstly
found in the precision measurements involving top-quark. A
possible signature for new physics could be demonstrated in the
deviation of the any of the $tt\gamma$, $ttZ$ and $tbW$ couplings
from the predictions of the SM. Until now there have been many
works which devote to the effects of new physics on the
observables related to the top-quark
couplings\cite{Berger,examples}. All these studies indicate that
the vector and axial form factors in the couplings of top-quarks
and gauge boson could receive large corrections in certain models
of dynamical EW breaking\cite{ttz}.

\par
As we know that it is very hard to obtain the information about
the $t\bar tV$ $(V=Z^0, \gamma)$ coupling from the precise
measurement of the top-pair production $e^+e^-\to\gamma^*/Z^*\to
t\bar{t}$ at a linear collider, because of the difficulty in
distinguishing the contributions of $t\bar tZ^0$ and $t\bar
t\gamma$ couplings. The probing of $t\bar t \gamma$ coupling via
$\gamma\gamma \to t\bar t$ process was investigated in
Ref.\cite{ggtt}. It is found that the $\gamma\gamma \to t\bar t$
process is sensitive to $t\bar t \gamma$ coupling. In references
\cite{a7,han,zhou}, the QCD and EW corrections to top quark pair
production via fusion of both polarized and unpolarized photons in
the minimal supersymmetric standard model(MSSM), are presented.
The corrections are found to be sizable. At hadron colliders, a
measurement of the EW neutral couplings via
$q\bar{q}\to\gamma^*/Z^*\to t\bar{t}$ is hopeless due to the
strong interaction process $q\bar{q}(gg)\to g^*\to t\bar{t}$.
Instead, they can be measured in QCD $t\bar{t}Z^0/\gamma$
production and radiative top quark decays in $t\bar{t}$ events
($t\bar{t}\to\gamma W^+W^- b\bar{b}$). Each of the processes is
sensitive to the EW couplings between top-quark and the emitted
$Z^0$-boson(or photon)\cite{Baur}. The coupling of top-quark and
$Z^0$ gauge boson can be probed via the measurement of process
\eettz at linear colliders. The process \eettz at the tree-level
has been already discussed in the literature \cite{Hagiwara} by
Hagiwara in the context of the SM, while CP-violating effects in
\eettz process were studied in the framework of the Two Higgs
Doublets Model\cite{CPT-THDM} and with model independent effective
Lagrangian\cite{CPT-Lag}. Recently, the NLO QCD corrections to
$t\bar t Z^0$ production at the LHC have been calculated in
reference \cite{pp-ttz}.

\par
The International Linear Collider (ILC) is an efficient machine
for precise experiments designed in $e^+e^-$ colliding energy
range of $200~GeV<\sqrt{s}<500~GeV$, and it has an integrated
luminosity of around $500~(fb)^{-1}$ in four years. It would be
upgraded to $\sqrt{s}\sim 1~TeV$ with an integrated luminosity of
$1~(ab)^{-1}$ in three years\cite{ILC}. This machine has
sufficient energy to produce top-quarks, and be ideally suited to
precision studies of many top-quark properties due to the
advantage of the measurement being carried out in a particularly
clean environment. Thus the anomalous coupling between top-quarks
and $Z^0$-boson can be probed by measuring precisely the \eettz
process at a linear collider.

\par
In this work we present the calculations involving the full
one-loop QCD and EW corrections to the process \eettz in the SM.
The paper is arranged as follows: In Section II we give the
analytical calculation description of the Born cross section. The
calculations of full ${\cal O}(\alpha_{s})$ QCD and ${\cal
O}(\alpha_{ew})$ electroweak radiative corrections to
$t\bar{t}Z^0$ production at an $e^+e^-$ linear collider are
provided in Section III and IV, respectively. In Section V we
present some numerical results and discussions, and finally a
short summary is given.

\vskip 10mm
\section{LO calculation for \eettz process}
\par
\begin{figure*}
\begin{center}
\includegraphics*[120pt,500pt][530pt,670pt]{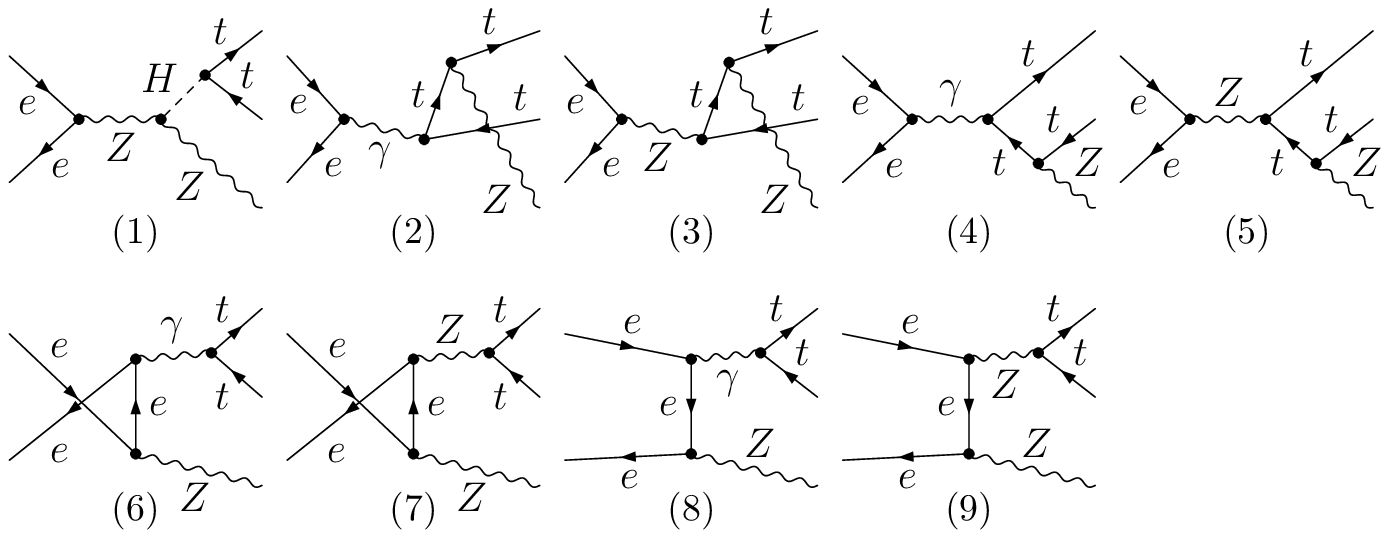}
\caption{The lowest order diagrams for the \eettz process in the
SM.} \label{fig1}
\end{center}
\end{figure*}

We use the 't Hooft-Feynman gauge in the leading-order
calculation, except when we verify the gauge invariance. The
contribution to the cross section of process \eettz in the SM at
the lowest order is of the order ${\cal O}(\alpha_{ew}^3)$ with
pure electroweak interactions. There are totally nine tree-level
Feynman diagrams, which are drawn in Fig.\ref{fig1}. The Feynman
diagrams in Fig.\ref{fig1} topologically can be divided into
s-channel(Fig.\ref{fig1}(1)-(5)) and
t(u)-channel(Fig.\ref{fig1}(6)-(9)) diagrams.

\par
We define the notations for the process \eettz as
\begin{equation}
e^+(p_1)+e^-(p_2) \to t(k_1)+\bar{t}(k_2)+Z^0(k_3),
\end{equation}
where the four-momenta of incoming positron and electron are
represented as $p_{1}$ and $p_{2}$, respectively, and the
four-momenta of outgoing top-quark, anti-top-quark and $Z^0$-boson
are denoted as $k_1$, $k_2$ and $k_3$ correspondingly. All these
momenta obey the on-shell equations $p_1^2=p_2^2=m_e^2$,
$k_1^2=k_2^2=m_t^2$ and $k_3^2=m_Z^2$.

\par
The amplitudes of the corresponding tree-level Feynman diagrams
for the process \eettz shown in Fig.\ref{fig1}, are respectively
expressed as
\begin{eqnarray}\label{eq1}
{\cal M}^{(1)}&=& \frac{ie^3m_t}{8s_W^3c_W^3} \times
\frac{1}{\left(s-m_Z^2\right)\left[(k_1+k_2)^2-m_H^2\right]}\bar{u}(k_1)v(k_2)
\bar{v}(p_1)\gamma_{\mu}\left(1-4s_W^2-\gamma_5\right)~~~~\nb \\
&\times&  u(p_2)\epsilon^{\mu}(k_3),
\end{eqnarray}
\begin{eqnarray}\label{eq2}
{\cal M}^{(2)}&=& \frac{-ie^3}{6s_W c_W} \times
\frac{1}{s\left[(k_1+k_3)^2-m_t^2\right]}\bar{u}(k_1)
\gamma_{\mu}\left(1-\frac{8}{3}s_W^2-\gamma_5\right)
(\rlap/k_1+\rlap/k_3+m_t) ~~~~\nb \\
&\times&
\gamma_{\nu}v(k_2)\bar{v}(p_1)\gamma^{\nu}u(p_2)\epsilon^{\mu}(k_3),
\end{eqnarray}
\begin{eqnarray}\label{eq3}
{\cal M}^{(3)}&=& \frac{-ie^3}{64s_W^3 c_W^3} \times
\frac{1}{\left(s-m_Z^2\right)\left[(k_1+k_3)^2-m_t^2\right]}\bar{u}(k_1)
\gamma_{\mu}\left(1-\frac{8}{3}s_W^2-\gamma_5\right)
(\rlap/k_1+\rlap/k_3+m_t) ~~~~\nb \\
&\times&\gamma^{\nu}\left(1-\frac{8}{3}s_W^2-\gamma_5\right)v(k_2)
\bar{v}(p_1)\gamma_{\nu}\left(1-4s_W^2-\gamma_5\right)u(p_2)\epsilon^{\mu}(k_3),
\end{eqnarray}
\begin{eqnarray}\label{eq4}
{\cal M}^{(4)}&=& \frac{-ie^3}{6s_W c_W} \times
\frac{1}{s\left[(k_2+k_3)^2-m_t^2\right]}\bar{u}(k_1)\gamma^{\nu}
(-\rlap/k_2-\rlap/k_3+m_t)\gamma_{\mu}
\left(1-\frac{8}{3}s_W^2-\gamma_5\right) ~~~~\nb \\
&\times&  v(k_2)
\bar{v}(p_1)\gamma_{\nu}u(p_2)\epsilon^{\mu}(k_3),
\end{eqnarray}
\begin{eqnarray}\label{eq5}
{\cal M}^{(5)}&=& \frac{-ie^3}{64s_W^3 c_W^3} \times
\frac{1}{\left(s-m_Z^2\right)\left[(k_2+k_3)^2-m_t^2\right]}
\bar{u}(k_1)\gamma^{\nu}\left(1-\frac{8}{3}s_W^2-\gamma_5\right)
(-\rlap/k_2-\rlap/k_3+m_t) \gamma_{\mu}~~~~\nb \\
 &\times&  \left(1-\frac{8}{3}s_W^2-\gamma_5\right) v(k_2)
\bar{v}(p_1) \gamma_{\nu} \left(1-4s_W^2-\gamma_5\right)u(p_2)
\epsilon^{\mu}(k_3),
\end{eqnarray}
\begin{eqnarray}\label{eq6}
{\cal M}^{(6)}&=& \frac{ie^3}{6s_W c_W} \times
\frac{1}{(k_1+k_2)^2\left[(p_2-k_3)^2-m_e^2\right]} \bar{u}(k_1)
\gamma^{\nu}v(k_2)
\bar{v}(p_1) \gamma_{\nu}(\rlap/p_2-\rlap/k_3+m_e)~~~~\nb \\
&\times&  \gamma_{\mu} \left(1-4s_W^2-\gamma_5\right)u(p_2)
\epsilon^{\mu}(k_3),
\end{eqnarray}
\begin{eqnarray}\label{eq7}
{\cal M}^{(7)}&=& \frac{ie^3}{64s_W^3 c_W^3} \times
\frac{1}{\left[(k_1+k_2)^2-m_Z^2\right]\left[(p_2-k_3)^2-m_e^2\right]}
\bar{u}(k_1)\gamma^{\nu}\left(1-\frac{8}{3}s_W^2-\gamma_5\right)v(k_2)
\bar{v}(p_1)\gamma_{\nu}~~~~\nb \\
&\times&  \left(1-4s_W^2-\gamma_5\right)
(\rlap/p_2-\rlap/k_3+m_e)\gamma_{\mu}
\left(1-4s_W^2-\gamma_5\right) u(p_2)\epsilon^{\mu}(k_3),
\end{eqnarray}
\begin{eqnarray}\label{eq8}
{\cal M}^{(8)}&=& \frac{ie^3}{6s_W c_W} \times
\frac{1}{(k_1+k_2)^2\left[(p_1-k_3)^2-m_e^2\right]}\bar{u}(k_1)
\gamma^{\nu} v(k_2) \bar{v}(p_1) \gamma_{\mu}
(-\rlap/p_1+\rlap/k_3+m_e) ~~~~\nb \\
&\times&  \gamma_{\nu}\left(1-4s_W^2-\gamma_5\right)
u(p_2) \epsilon^{\mu}(k_3),
\end{eqnarray}
\begin{eqnarray}\label{eq9}
{\cal M}^{(9)}&=& \frac{ie^3}{64s_W^3 c_W^3} \times
\frac{1}{\left[(k_1+k_2)^2-m_Z^2\right]\left[(p_1-k_3)^2-m_e^2\right]}
\bar{u}(k_1)\gamma^{\nu}\left(1-\frac{8}{3}s_W^2-\gamma_5\right)
v(k_2)\bar{v}(p_1)~~~~\nb \\
&\times& \gamma_{\mu} \left(1-4s_W^2-\gamma_5\right)
(-\rlap/p_1+\rlap/k_3+m_e)\gamma_{\nu} \left(1-4s_W^2-
\gamma_5\right)u(p_2) \epsilon^{\mu}(k_3).
\end{eqnarray}

\par
Then the total amplitude at the lowest order can be obtained by
summing up all the above amplitudes given in
Eqs.(\ref{eq1})-(\ref{eq9}).
\begin{equation}
{\cal M}_0=\sum_{i=1}^{9} {\cal M}^{(i)}.
\end{equation}

\par
The differential cross section for the process \eettz at the
tree-level with unpolarized incoming particles is then obtained as
\begin{eqnarray} \label{cross}
d\sigma_{0} =\frac{N_c}{4}\sum_{spins}|{\cal M}_0|^2 d\Phi_3 ,
\end{eqnarray}
where the color number $N_c=3$, the summation is taken over the
spins of initial and final particles. The factor $\frac{1}{4}$ is
due to taking average over the polarization states of the electron
and positron. $d\Phi_3$ is the three-particle phase space element
defined as
\begin{eqnarray}
d\Phi_3=\delta^{(4)} \left( p_1+p_2-\sum_{i=1}^3 k_i \right)
\prod_{j=1}^3 \frac{d^3 \textbf{\textsl{k}}_j}{(2 \pi)^3 2 E_j}.
\end{eqnarray}

\vskip 10mm
\section{ One-loop QCD corrections to the \eettz process}
\begin{figure*}
\begin{center}
\includegraphics*[120pt,260pt][530pt,660pt]{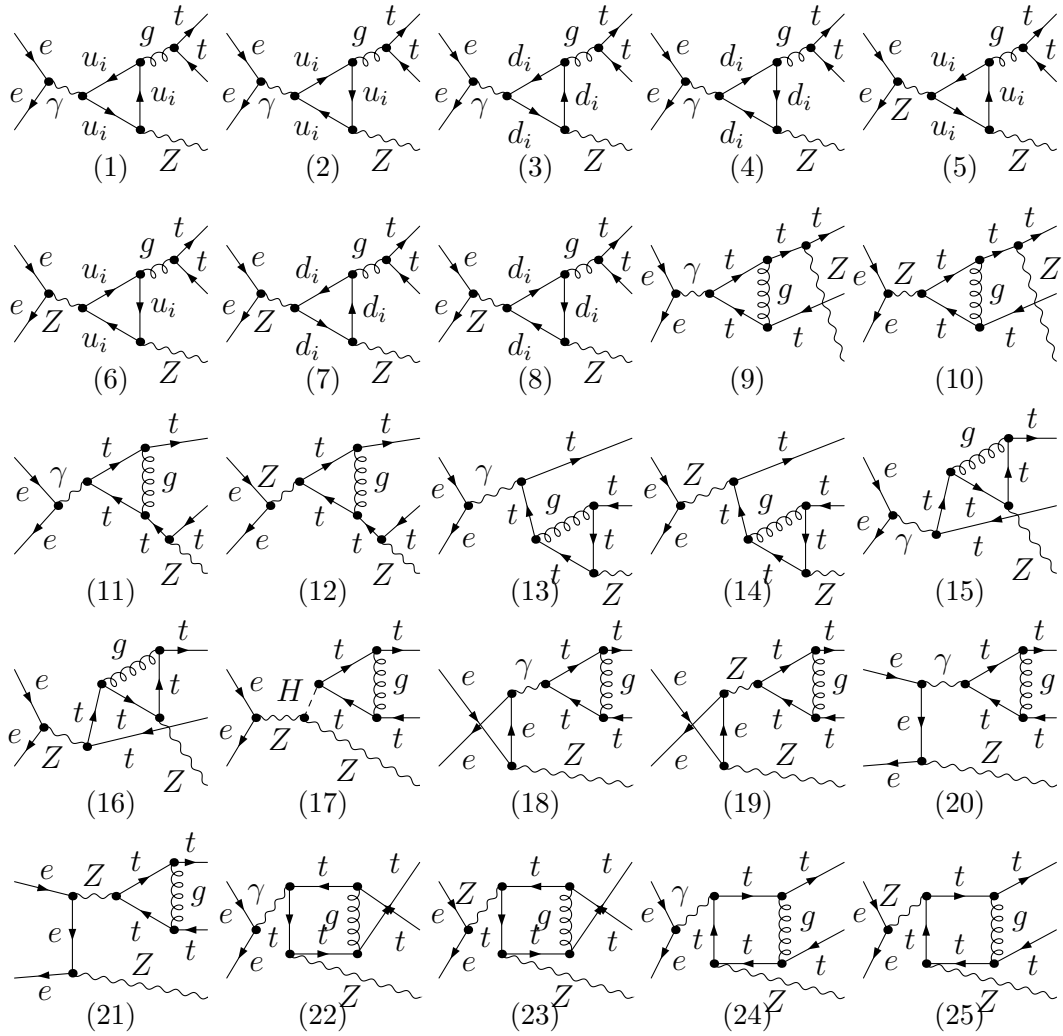}
\caption{The box and triangle diagrams of one-loop QCD corrections
to the \eettz process. } \label{fig2a}
\end{center}
\end{figure*}

\par
The QCD one-loop Feynman diagrams and corresponding amplitudes of
the process \eettz are generated by using FeynArts3.2 \cite{fey}. We
present the box and triangle diagrams of one-loop QCD corrections to
the \eettz process in Fig.\ref{fig2a}. We take the definitions of
one-loop integral functions in Ref.\cite{OMS}, and use mainly the
FormCalc4.1 package\cite{formloop} to calculate the amplitudes of
one-loop Feynman diagrams and get the results in a way well suited
for further numerical or analytical evaluation. The calculation of
the amplitude for one-loop diagram has been performed in the
conventional 't Hooft--Feynman gauge by adopting dimension
regularization scheme in which the dimensions of spinor and
space-time manifolds are extended to $D = 4 - 2 \epsilon$. The
numerical calculations of the integral functions are implemented by
using our in-house programs developed from the FF package\cite{ff}.
In these programs, the numerical evaluations of one-point,
two-point, three-point and four-point integrals are evaluated by
using the expressions from Ref.\cite{Velt}, and the scalar and
tensor 5-point integrals are calculated exactly by using the
approach presented in Ref.\cite{Five}. The phase space integration
for the process \eettz is implemented by adopting 2to3.F subroutine
in FormCalc4.1 package. In the calculation of the correction from
the hard gluon emission process \eettzg, we use CompHEP-4.4p3
program\cite{CompHEP} to generate all the tree-level diagrams and
their corresponding amplitudes automatically, and accomplish the
four-body phase-space integration. In CompHEP-4.4p3 and 2to3.F
programs the well-known adaptive Monte Carlo method (the VEGAS
algorithm) is adopted for multi-particle phase space
integration\cite{Vegas}. Then we can obtain the NLO QCD corrected
total and differential cross sections. The amplitude of the process
\eettz including virtual QCD corrections at ${\cal O}(\alpha_s)$
order can be expressed as
\begin{equation}
{\cal M}_{QCD}={\cal M}_0+ {\cal M}_{QCD}^{vir},
\end{equation}
where ${\cal M}_{QCD}^{vir}$ is the renormalized amplitude
contributed by the QCD one-loop Feynman diagrams and their
corresponding counterterms. The relevant renormalization constants
for top field are defined as
\begin{eqnarray}
t_0^{L}=(1+\frac{1}{2}\delta Z_{t,{QCD}}^L)t^L, ~~
t_0^{R}=(1+\frac{1}{2}\delta Z_{t,{QCD}}^R)t^R.
\end{eqnarray}
Taking the on-mass-shell renormalized condition we get the
${\cal O}(\alpha_{s})$ QCD contributed parts of the renormalization
constants as
\begin{eqnarray}
\label{counterterm 1} \delta Z_{t,{QCD}}^{L} = -
\tilde{Re}\Sigma_{t,{QCD}}^{L}(m_{t}^{2})
    - m_{t}^2 \frac{\partial}{\partial p^2}
        \tilde{Re}\left [ \Sigma_{t,{QCD}}^{L}(p^2) +
      \Sigma_{t,{QCD}}^{R}(p^2) + 2\Sigma_{t,{QCD}}^{S}(p^2) \right ] |_{p^2=m_{t}^2},
\end{eqnarray}
\begin{eqnarray}
\label{counterterm 2} \delta Z_{t,{QCD}}^{R} =
-\tilde{Re}\Sigma_{t,{QCD}}^{R}(m_{t}^{2}) -
    m_{t}^2 \frac{\partial}{\partial p^2}
        \tilde{Re}\left [ \Sigma_{t,{QCD}}^{L}(p^2) +
                \Sigma_{t,{QCD}}^{R}(p^2) +
                2 \Sigma_{t,{QCD}}^{S}(p^2)  \right ] |_{p^2=m_{t}^2},
\end{eqnarray}
where $\tilde{Re}$ represents taking the real part of the loop
integrals appearing in the self-energies only, and the
renormalized top-quark irreducible QCD two-point function is defined
as
\begin{eqnarray}
\hat{\Gamma}_{t,{QCD}}(p^2)=i \left[ \rlap/p P_L
\hat{\Sigma}_{t,{QCD}}^{L}(p^2) + \rlap/p P_R
\hat{\Sigma}_{t,{QCD}}^{R}(p^2)+ m_{t}
\hat{\Sigma}_{t,{QCD}}^{S}(p^2)\right] \delta_{\alpha \beta}
\end{eqnarray}
where $\alpha$ and $\beta$ are the color indices of the top-quarks
on the two sides of the self-energy diagram, and
$P_{L,R}=(1\mp\gamma_5)/2$. The unrenormalized top-quark QCD
self-energy parts at QCD one-loop order read
\begin{eqnarray}
\Sigma_{t,{QCD}}^{L}(p^2)=\Sigma_{t,{QCD}}^{R}(p^2)=\frac{g_s^2C_F}{16
\pi^2}\left( -1+2B_0[p,0,m_t]+2B_1[p,0,m_t] \right),
\end{eqnarray}
and
\begin{eqnarray}
\Sigma_{t,{QCD}}^{S}(p^2)=\frac{g_s^2C_F}{8
\pi^2}\left(1-2B_0[p,0,m_t]\right).
\end{eqnarray}
The corresponding contribution part to the cross section at ${\cal
O}(\alpha_{s})$ order can be written as
\begin{eqnarray}
\Delta \sigma^{QCD}_{vir} = \sigma_{0} \delta^{QCD}_{vir} =
\frac{(2\pi)^4 N_c}{2|\vec{p}_1| \sqrt{s}} \int {\rm d} \Phi_3
\overline{\sum_{spins}} {\rm Re} \left( {\cal M}_{0}^{\dag} {\cal
M}_{QCD}^{vir} \right),
\end{eqnarray}
where $N_c=3$, $C_F=4/3$, and the bar over summation recalls
averaging over the spins of incoming particles.

\par
The virtual QCD correction contains ultraviolet (UV) and infrared
(IR) divergences. The UV divergences from one-loop diagrams can be
cancelled after the renormalization by adopting the dimensional
regularization. We have verified the UV finiteness for the
renormalized amplitude analytically and numerically.

\par
The IR divergences in the ${\cal M}_{QCD}^{vir}$ are originated
from the contributions of virtual gluon exchange in loops. These
IR divergencies can be cancelled by the real gluon bremsstrahlung
corrections in the soft gluon limit, due to the
Kinoshita-Lee-Nauenberg (KLN) theorem\cite{KLN}. In our
calculation these IR divergencies are regulated by a small gluon
mass. The real gluon emission process is denoted as
\begin{eqnarray}
\label{real g emission}
 e^+(p_1)+e^-(p_2) \to t(k_1)+\bar{t}(k_2)+Z^0(k_3)+g(k),
\end{eqnarray}
where the real gluon radiates from top(anti-top) quark line. We
adopt the phase-space-slicing method \cite{PSS} to isolate the IR
soft singularity. The cross section of the real gluon emission
process ($\ref{real g emission}$) is decomposed into soft and hard
terms
\begin{equation}
\Delta \sigma_{real}^{QCD}=\Delta \sigma_{ soft}^{QCD}+\Delta
\sigma_{hard}^{QCD}.
\end{equation}
In practice, we take a very small value for the cutoff $\Delta
E_g/E_{b}$ in the numerical calculations, and neglect the terms of
order $\Delta E_g/E_{b}$ in evaluating the following integration
for soft contribution\cite{OMS,Velt,ee1},
\begin{eqnarray}
\label{s} {d} \Delta\sigma_{soft}^{QCD} = - d \sigma_{{0}}
\frac{\alpha_{s}C_F}{2 \pi^2} \int_{E_{g} \leq \Delta E_{g}}
\frac{d^3 k}{2 E_{g}} \left( \frac{k_1}{k_1 \cdot k} - \frac{k_2}{k_2
\cdot k} \right)^2,
\end{eqnarray}
where $E_g \leq \Delta E_g \ll \sqrt{s}$, $\Delta E_g$ is the
energy cutoff of the soft gluon,  and the gluon energy can be
obtained by $E_g = \sqrt{|\vec{k}|^2+m_g^2}$.

\par
The cancelation of the IR divergencies in the virtual contribution
part and the soft gluon bremsstrahlung correction, can be verified
during our calculation. Thus we get an IR finite cross section which
is independent of the infinitesimal fictitious gluon mass $m_g$. The
hard gluon emission cross section $\Delta \sigma^{QCD}_{hard}$ is
calculated numerically by using Monte Carlo program in CompHEP-4.4p3
package.

\par
The UV and IR finite total cross section of the process \eettz
including the ${\cal O}(\alpha_{s})$ QCD contributions reads
\begin{equation}\label{cs}
\sigma^{QCD}  = \sigma_0 + \Delta \sigma^{QCD} = \sigma_0 +
\Delta\sigma_{vir}^{QCD} + \Delta\sigma_{soft}^{QCD} +
\Delta\sigma_{hard}^{QCD} = \sigma_0(1 + \delta^{QCD}),
\end{equation}
where $\Delta\sigma^{QCD}$ and $\delta^{QCD}$ are the QCD cross
section correction and QCD relative correction at the order of
${\cal O}(\alpha_s)$, respectively.

\vskip 10mm
\section{ One-loop EW corrections to the \eettz process}
\par
\begin{figure*}
\begin{center}
\includegraphics*[120pt,500pt][530pt,660pt]{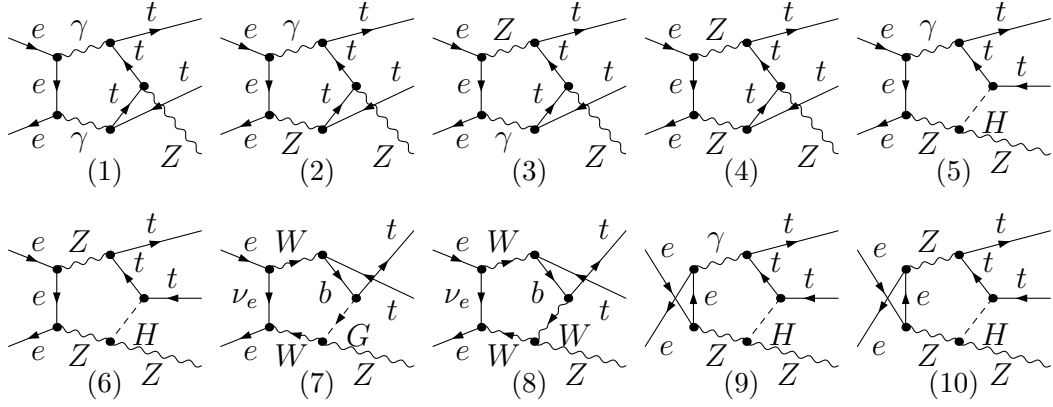}
\caption{Some representative pentagon one-loop EW diagrams for the
\eettz process. } \label{fig2b}
\end{center}
\end{figure*}

In this section, we present the calculation of the ${\cal
O}(\alpha_{ew})$ order EW corrections to the process \eettz. We
use again the package FeynArts3.2 to generate the one-loop EW
Feynman diagrams and the relevant amplitudes of the process
\eettz. The one-loop EW Feynman diagrams can be classified into
self-energy, triangle, box, pentagon and counterterm diagrams.
Some pentagon diagrams are depicted in Fig.\ref{fig2b} as a
representative selection. In our EW correction calculation we
adopt the t'Hooft-Feynman gauge and the same definitions of
one-loop integral functions as in Ref.\cite{OMS}. We take the
dimensional regularization scheme to regularize the UV divergences
in loop integrals, and assume that the CKM matrix is identity
matrix and use the on-mass-shell(OMS) renormalization scheme
\cite{OMS}. The relevant field and mass renormalization constants
are defined as
\begin{eqnarray}
e_0=(1+\delta Z_e)e,~~~m_{f,0}=m_f+\delta m_f,~~~
m_{W(Z),0}^2=m_{W(Z)}^2+\delta m_{W(Z)}^2, ~~~H_0=(1+\frac{1}{2} \delta Z_H)H,~~~ \nb \\
A_0=\frac{1}{2}\delta Z_{AZ}Z+(1+\frac{1}{2}\delta Z_{AA})A,~~~
Z_0=(1+\frac{1}{2}\delta Z_{ZZ})Z+\frac{1}{2}\delta Z_{ZA}A,~~~~~~~~~~~~  \nb \\
f_0^{L}=(1+\frac{1}{2}\delta Z_f^L)f^L,  ~~~~
f_0^{R}=(1+\frac{1}{2}\delta Z_f^R)f^R.~~~~~~~~~~~~~~~
\end{eqnarray}
With the on-mass-shell conditions, we can obtain the related
renormalized constants expressed as
\begin{eqnarray}
\delta m_f=\frac{m_f}{2}\tilde{Re}\left[ \Sigma_f^L(m_f^2)+
\Sigma_f^R(m_f^2)+2\Sigma_f^S(m_f^2) \right],~~~~~~~~~\nb \\
\delta m_{W}^2=\tilde{Re}\Sigma_T^{W}(m_{W}^2),~~~\delta
m_{Z}^2=Re\Sigma_T^{ZZ}(m_{Z}^2),~~~\delta Z_H=-Re\frac{\partial
\Sigma^H(k^2)}{\partial
k^2}|_{k^2=m_H^2},~~~\nb \\
\delta Z_{AA}=- \frac{\partial \Sigma_T^{AA}(p^2)}{\partial
p^2}|_{p^2=0},~~~ \delta Z_{ZZ}= - Re \frac{\partial
\Sigma_T^{ZZ}(p^2)}{\partial
p^2}|_{p^2=m_Z^2}, ~~~ \nb \\
\delta Z_{ZA}=2\frac{\Sigma_T^{ZA}(0)}{m_Z^2}, ~~~ \delta Z_{AZ}=
- 2 Re \frac{\Sigma_T^{AZ}(m_Z^2)}{m_Z^2}.
\end{eqnarray}
The charge renormalization constant and the counterterm of the
parameter $s_W$ can be obtained by using following
equations\cite{OMS}:
\begin{eqnarray}
\delta Z_e=-\frac{1}{2}\delta
Z_{AA}-\frac{s_W}{c_W}\frac{1}{2}\delta Z_{ZA},~~~\frac{\delta
s_W}{s_W}=-\frac{1}{2}\frac{c_W^2}{s_W^2}\tilde{Re}\left(
\frac{\Sigma_T^W(m_W^2)}{m_W^2}-\frac{\Sigma_T^{ZZ}(m_Z^2)}{m_Z^2}\right).
\end{eqnarray}

\par
We use Eqs.(\ref{counterterm 1})-(\ref{counterterm 2}) to
calculate the renormalization constants of the fermion wave
functions, but the top-quark QCD self-energies in these
equations($\Sigma_{t,{QCD}}^{L}$, $\Sigma_{t,{QCD}}^{R}$ and
$\Sigma_{t,{QCD}}^{S}$) are replaced by the corresponding fermion
EW self-energies($\Sigma_{f,EW}^{L}$, $\Sigma_{f,EW}^{R}$ and
$\Sigma_{f,EW}^{S}$), respectively. And the explicit expressions
of the relevant EW self-energies in the SM can be found in the
Appendix B of Ref.\cite{OMS}. The UV divergence appearing from the
one-loop diagrams should be cancelled by the contributions of the
counterterm diagrams. In our calculations, it was verified both
analytically and numerically that the total cross sections
including ${\cal O}(\alpha_{ew})$ one-loop radiative corrections
and the corresponding counterterm contributions, are UV finite.
Then the one-loop level virtual EW corrections to the cross
section can be expressed as
\begin{eqnarray}
\Delta \sigma^{EW}_{{\rm vir}} = \sigma_{0} \delta^{EW}_{vir} =
\frac{(2\pi)^4 N_c}{2 |\vec{p}_1| \sqrt{s}} \int {\rm d} \Phi_3
\overline{\sum_{spins}} {\rm Re} \left( {\cal M}^{\dag}_{0}
{\cal M}^{EW}_{{\rm vir}} \right),
\end{eqnarray}
where $\vec{p}_1$ is the c.m.s. three-momentum of the incoming
positron, ${\rm d} \Phi_3$ is the three-body phase space element,
and the bar over summation denotes averaging over initial spins.
${\cal M}^{EW}_{{\rm vir}}$ is the renormalized total amplitude of
all the one-loop EW level Feynman diagrams, including self-energy,
vertex, box, pentagon and counterterm diagrams.

\par
Analogous to the calculation of the QCD correction as described in
above section, we generate the EW one-loop Feynman diagrams and
calculate their amplitudes for \eettz process by using FeynArts3.2
and FormCalc4.1 packages. The numerical calculations of integral
functions are performed by adopting our in-house library. The
phase space of process \eettz is integrated by using 2to3.F
subroutine. In calculating the correction from hard photon
emission process \eettzga, the CompHEP-4.4p3 program is used to
generate the tree-level diagrams and amplitudes, and carry out the
integration in four-body phase-space. During the calculation of
the EW corrections to the process \eettz, the IR divergence
originating from virtual photon correction should be canceled by
the real photon bremsstrahlung correction in the soft photon
limit. We use also the phase-space-slicing method and divide the
cross section of the real photon emission process, denoted as
$e^+(p_1)e^-(p_2) \to t(k_1) \bar{t}(k_2) Z^0(k_3) \gamma(k)$,
into soft and hard parts.
\begin{equation}
\Delta \sigma_{{\rm real}}^{EW}=\Delta \sigma_{soft}^{EW}+\Delta
\sigma_{hard}^{EW}= \sigma_0( \delta^{EW}_{soft}+ \delta^{EW}_{
hard}).
\end{equation}
By using the soft photon($E_{\gamma} <\Delta E_{\gamma}$) approximation, we get the
contribution of the soft photon emission process expressed as
\begin{eqnarray}
\label{approsoft} {\rm d} \Delta \sigma^{EW}_{soft} = -{\rm d}
\sigma_0 \frac{\alpha_{ew}}{2 \pi^2}
 \int_{E_{\gamma} \leq \Delta E_{\gamma}}\frac{{\rm d}^3 k}{2 E_{\gamma}} \left[
 \frac{Q_t k_1}{k_1\cdot k}-\frac{Q_t k_2}{k_2\cdot k} -
 \frac{p_1}{p_1\cdot k} + \frac{p_2}{p_2\cdot k} \right]^2 ,
\end{eqnarray}
in which $\Delta E_{\gamma}$ is the energy cutoff of the soft
photon and $E_{\gamma} \leq \Delta E_{\gamma} \ll \sqrt{s}$,
$Q_t=2/3$ is the electric charge of top quark, $E_{\gamma} =
\sqrt{|\vec{k}|^2+m_{\gamma}^2}$ is the photon energy. Therefore,
after integrating approximately over the soft photon phase space,
we can obtain the analytical result of the soft photon emission
corrections to \eettz. The cancellation of IR divergencies is
verified and the results shows that the cross section for $e^+e^-
\to t\bar{t}Z^0g$ is independent on the infinitesimal photon mass
$m_{\gamma}$ in our calculation.

\par
For the convenience in analyzing the origins of the EW radiation
corrections, we split the full one-loop EW level correction into
the QED correction part ($\Delta \sigma^{E}$) and the pure weak
correction part ($\Delta \sigma^{W}$). The QED correction part
involves the contributions from the one-loop diagrams with virtual
photon exchange in loop, the real photon emission process
\eettzga, and the pure photonic contribution of the
renormalization constants of the related fermions. The rest of the
total EW corrections remains with the weak correction part. Then
we can express the full one-loop EW corrected total cross section
as
\begin{eqnarray}
\Delta \sigma^{EW}= \Delta \sigma^{E} + \Delta \sigma^{W}.
\end{eqnarray}

\vskip 10mm
\section{ Numerical Results and Discussion}
\par
In the following numerical calculation, we have used the 3-loop
evolution of strong coupling constant $\alpha_s(\mu^2)$ in the
$\overline{MS}$ scheme with parameters
$\Lambda_{QCD}^{n_f=5}=203.73~MeV$, yielding
$\alpha_{s}^{\overline{MS}}(m_Z^2)=0.1176$. We take the relevant
parameters as\cite{tew,hepdata} : $\alpha_{{\rm ew}}(0)^{-1} =
137.036$, $m_e=0.519991~MeV$, $m_{\mu}=105.6583692~MeV$,
$m_{\tau}=1.77699~GeV$, $m_W = 80.403~GeV$, $m_Z = 91.1876~GeV$,
$m_t = 172.5~GeV$, $m_s=150~GeV$, $m_c=1.2~GeV$, $m_b=4.7~GeV$, $m_u
= m_d = 66~MeV$, $\sin^2 \theta_W=1-m_W^2/m_Z^2=0.222549$, and
$\mu=m_t+\frac{1}{2}m_Z$. There the light quark masses ($m_u$ and
$m_d$) have the effective values which can reproduce the hadron
contribution to the shift in the fine structure constant
$\alpha_{ew}(m_Z^2)$\cite{leger}. With above parameter choice we get
$\alpha_s(\mu^2=(m_t+\frac{1}{2}m_Z)^2)=0.1040$. Since LEP II and
the electroweak precision measurements have provided that the SM
Higgs-boson exists in the mass range of $114.4~ GeV < m_H \lesssim
182~GeV$\cite{lower mH,upper mH}, we include the relevant Higgs
contributions with $m_H =120~GeV$ as a representation.

\par
\begin{figure}
\includegraphics[scale=0.75]{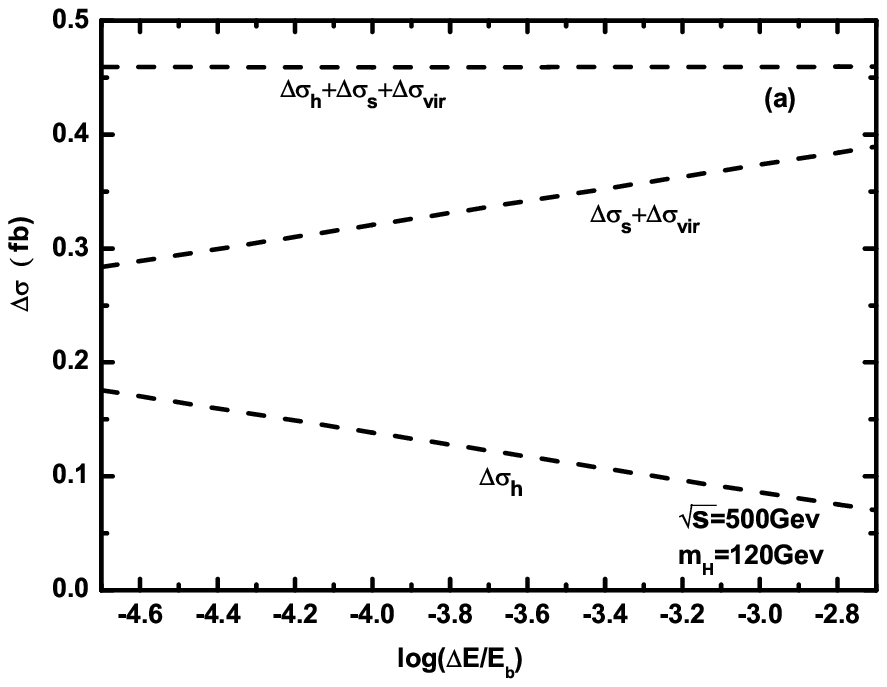}
\includegraphics[scale=0.75]{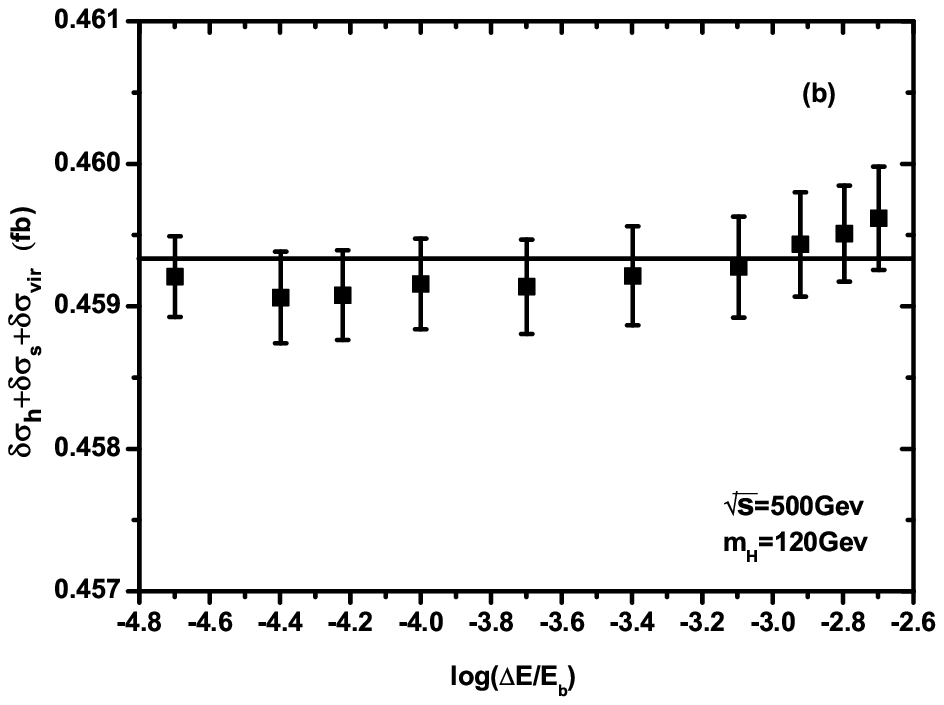}
\caption{\label{fig3} (a) The ${\cal O}(\alpha_{s})$ QCD corrections
to the cross section for \eettz as the functions of the soft cutoff
$\delta_s \equiv \Delta E_{g}/E_b$ in conditions of
$m_{g}=10^{-3}~GeV$, $m_H=120~GeV$ and $\sqrt{s}=500~GeV$. (b) The
amplified curve marked with the calculation errors for
$\Delta\sigma^{QCD}$ of Fig.\ref{fig3}(a) versus $\delta_s$.}
\end{figure}
\begin{figure}
\includegraphics[scale=0.75]{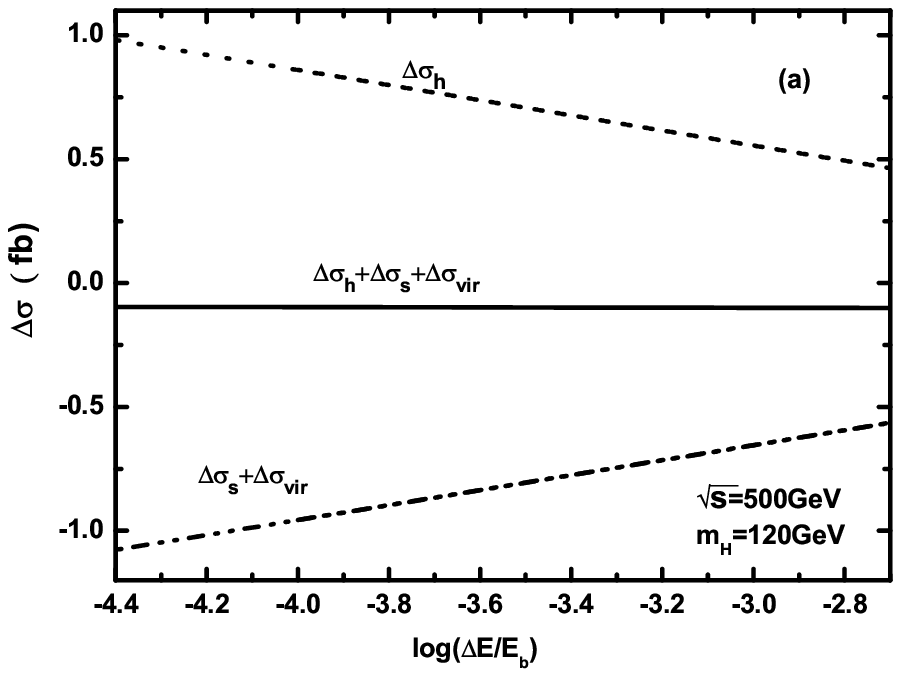}
\includegraphics[scale=0.75]{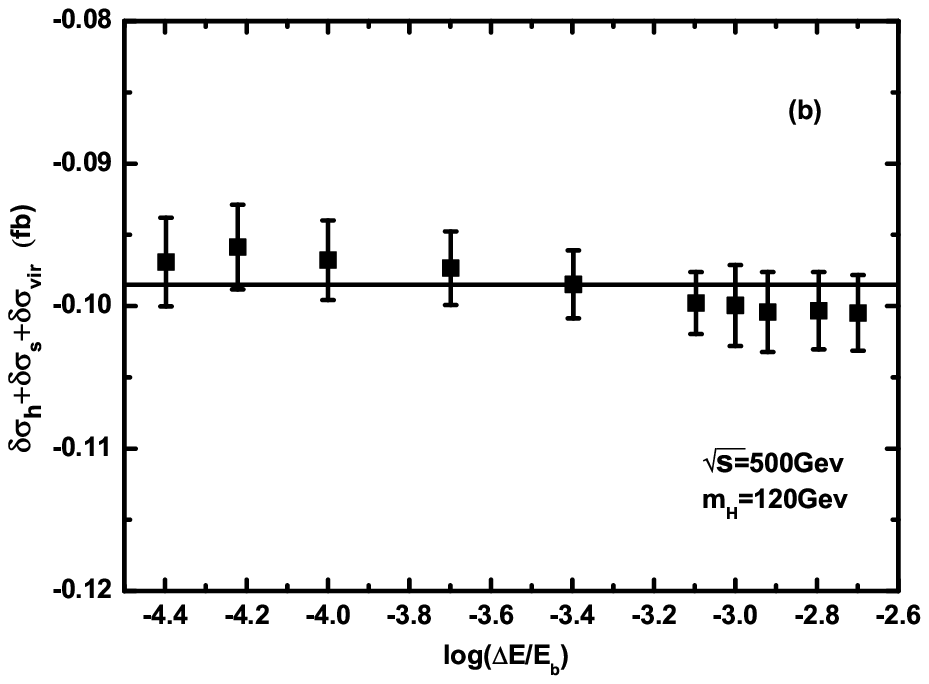}
\caption{\label{fig4} (a) The ${\cal O}(\alpha_{ew})$ EW corrections
to the cross section for \eettz as the functions of the soft cutoff
$\delta_s\equiv \Delta E_{\gamma}/E_b$ in conditions of
$m_{\gamma}=10^{-3}~GeV$, $m_H=120~GeV$ and $\sqrt{s}=500~GeV$. (b)
The amplified curve marked with the calculation errors for
$\Delta\sigma^{EW}$ of Fig.\ref{fig4}(a) versus $\delta_s$.}
\end{figure}

During our calculation of one-loop corrections, we have to fix the
values of the fictitious masses of photon and gluon(as IR
regulators $m_{\gamma}$ and $m_g$), and soft cutoff
$\delta_s=\Delta E_{g(\gamma)}/E_b$ except the input parameters
mentioned above. In fact, the physical total cross section should
be independent of these regulators and soft cutoff. We have
verified the invariance of the cross section contributions at QCD
and EW one-loop order,
$\Delta\sigma^{QCD,EW}_{tot}=\Delta\sigma_{real}^{QCD,EW}+\Delta\sigma_{vir}^{QCD,EW}$,
within the calculation errors when the fictitious photon and gluon
masses, $m_{\gamma}$ and $m_{g}$, vary from $10^{-10}~ GeV$ to
$10^{-1}~GeV$ in conditions of $\delta_s=2\times 10^{-3}$,
$m_H=120~GeV$ and $\sqrt{s}=500~GeV$. The relation between the
${\cal O}(\alpha_{s})$ QCD (${\cal O}(\alpha_{ew})$ EW) correction
and soft cutoff $\delta_s$ demonstrates in Figs.\ref{fig3}(a-b)
(Figs.\ref{fig4}(a-b)), assuming $m_{g}=10^{-3}~GeV$
($m_{\gamma}=10^{-3}~GeV$), $m_{H} = 120~GeV$ and $\sqrt{s} =
500~GeV$. We can see in Fig.\ref{fig3}(a) and Fig.\ref{fig4}(a)
that the curves for
$\Delta\sigma^{QCD,EW}_{soft}+\Delta\sigma^{QCD,EW}_{vir}$ and
$\Delta\sigma^{QCD,EW}_{hard}$ are strongly related to the soft
cutoff $\delta_s$, but both the total QCD and  EW one-loop
radiative corrections,
$\Delta\sigma^{QCD,EW}(\equiv\Delta\sigma^{QCD,EW}_{vir}+
\Delta\sigma^{QCD,EW}_{soft}+\Delta\sigma^{QCD,EW}_{hard})$, are
independent of the cutoff $\delta_s$ within the range of
calculation errors as we expected. In Fig.\ref{fig3}(b) and
Fig.\ref{fig4}(b), we make the curves for $\Delta\sigma^{QCD}$ and
$\Delta\sigma^{EW}$ greater in size and mark them with the
calculation errors, respectively. In the further calculation, we
fix the soft cutoff, fictitious photon and gluon masses as
$\delta_s=2\times 10^{-3}$, $m_{\gamma}=10^{-1}~GeV$ and
$m_{g}=10^{-2}~GeV$.

\par
In order to verify the reliability of our calculations for the
tree-level cross section of process \eettz, the QCD correction
from hard gluon emission process \eettzg and the EW correction
from hard photon emission process \eettzga, we evaluated their
numerical results by adopting different gauges and software tools.
In Table \ref{tab1}, we list these numerical results by taking
$m_H =120~GeV$ and $\sqrt{s}=500~GeV$, and using CompHEP-4.4p3
program\cite{CompHEP} (in both Feynman and unitary gauges),
FeynArts3.2/FormCalc4.1 \cite{fey,formloop}(in Feynman gauge, and
the phase space integrations are implemented by using 2to3.F
subroutine for \eettz process and in-house $2\to4$ phase space
integration program for \eettzg and \eettzga processes.) and
Grace2.2.1\cite{Grace} package(in Feynman gauge), separately. We
can see the results are in mutual agreement within the phase space
integration errors. We have calculated also the tree-level cross
section for the \eettz process by taking the same input parameters
as in Ref.\cite{Hagiwara}, and got coincident numerical results.
\begin{table}
\begin{tabular}{|c|c|c|c|c|}
\hline &    CompHEP   &  CompHEP    &    FeynArts   &  Grace    \\
\hline & Feynman Gauge & Unitary Gauge & Feynman Gauge     & Feynman Gauge \\
\hline $\sigma_0$ (fb) & 1.0649(1) & 1.0648(1) & 1.0648(2) & 1.0647(2) \\
\hline $\Delta\sigma_{hard}^{QCD}$ (fb) & $7.028(2)\times 10^{-2}$ &
$7.028(2)\times 10^{-2}$
& $7.028(2)\times 10^{-2}$ & $7.028(2)\times 10^{-2}$ \\
\hline $\Delta\sigma_{hard}^{EW}$ (fb) & $0.4631(2)$ & $0.4632(2)$
& $0.4631(2)$ & $0.4632(2)$ \\
\hline
\end{tabular}
\begin{center}
\begin{minipage}{15cm}
\caption{\label{tab1} The comparison of the results for the
tree-level cross section for the process \eettz($\sigma_0$) and the
corrections from hard gluon/photon emission processes
\eettzg($\Delta\sigma_{hard}^{QCD}$) and
\eettzga($\Delta\sigma_{hard}^{EW}$), in conditions of
$m_H=120~GeV$, $\sqrt{s}=500~GeV$ and $\delta_s=2\times 10^{-3}$.
The numerical results are obtained by using CompHEP-4.4p3(in both
Feynman and unitary gauges), FeynArts3.2/FormCalc4.1(in Feynman
gauge, the phase space integrations are implemented by using 2to3.F
subroutine for \eettz process and in-house $2\to4$ phase space
integration program for \eettzg and \eettzga processes) and
Grace2.2.1(in Feynman gauge) packages, separately. }
\end{minipage}
\end{center}
\end{table}

\par
For the check of the calculations for one-loop diagrams, we used the
LoopTools2.2 library and our in-house program(Both are in two
combination cases respectively) to calculate the one-loop QCD and EW
corrections to the process \eettz
($\Delta\sigma^{QCD,EW}_{soft}+\Delta\sigma^{QCD,EW}_{vir}$)
numerically, in conditions of $m_H=120~GeV$,
$\sqrt{s}=500~GeV,~800~GeV$, and other relevant parameters having
the values mentioned above. The LoopTools2.2 is used in two ways:
(I) with FF package(case I), (II) with routines adapted from A.
Denner's bcanew.f(case II). The in-house programs are also applied
in two ways: one is to use completely our created codes according to
the expressions in Refs.\cite{OMS,Five} for the numerical
calculations of N-point($N=1,2,3,4,5$) integrals (case I), another
way is to use our in-house program according to the formulas in
Ref.\cite{Five} for the numerical calculations of 5-point integrals
based on FF package(case II). In all cases we applied 2to3.F
subroutine in FormCalc4.1 package to perform the phase space
integrations. The numerical results are listed in Table \ref{tab3}.
Since among all the one-loop QCD Feynman diagrams for \eettz process
there is no pentagon diagram, we get the same results when we use
both in-house program of case (II) and LoopTools2.2 of case (I).
Table \ref{tab3} shows there exists the coincidence between the
corresponding results within the calculation errors.
\begin{table}
\begin{center}
\begin{tabular}{|c|c|c|c|c|c| }
\hline &    $\sqrt{s}(GeV)$   &  LoopTools2.2(I)  & LoopTools2.2(II)  & in-house(I)   & in-house(II)   \\
\hline $\Delta\sigma^{QCD}_{vir+soft}$ &    500  &  0.3893(3)& 0.3894(3) & 0.3892(3)  & 0.3893(3)   \\
\hline $\Delta\sigma^{QCD}_{vir+soft}$ &    800  & -1.396(1)
&-1.396(1)& -1.396(1)   & -1.396(1)   \\
\hline $\Delta\sigma^{EW}_{vir+soft}$ &    500  &  -0.5550(5)&
-0.5551(5) & -0.5550(5)   & -0.5551(5) \\
\hline $\Delta\sigma^{EW}_{vir+soft}$ &    800  &  -2.996(3) & -2.996(3) & -2.995(3)  & -2.996(3) \\
\hline
\end{tabular}
\end{center}
\begin{center}
\begin{minipage}{15cm}
\caption{\label{tab3} The comparison of the results for the QCD and
EW one-loop corrections to the process
\eettz($\Delta\sigma^{QCD}_{soft}+\Delta\sigma^{QCD}_{vir}$ and
$\Delta\sigma^{EW}_{soft}+\Delta\sigma^{EW}_{vir}$ ) numerically, in
conditions of $m_H=120~GeV$, $\sqrt{s}=500~GeV,~800~GeV$, and other
relevant parameters having the values mentioned above. The numerical
results are obtained by using LoopTools2.2 library in two
combinations((I) with FF package, (II) with routines adapted from A.
Denner's bcanew.f), our developed in-house codes for all the
numerical calculations of N-point($N=1,2,3,4,5$) integrals(case I),
and our developed program for the numerical calculations of
five-point integrals based on FF package(case II). }
\end{minipage}
\end{center}
\end{table}

\par
\begin{figure}
\centering
\includegraphics[scale=0.75]{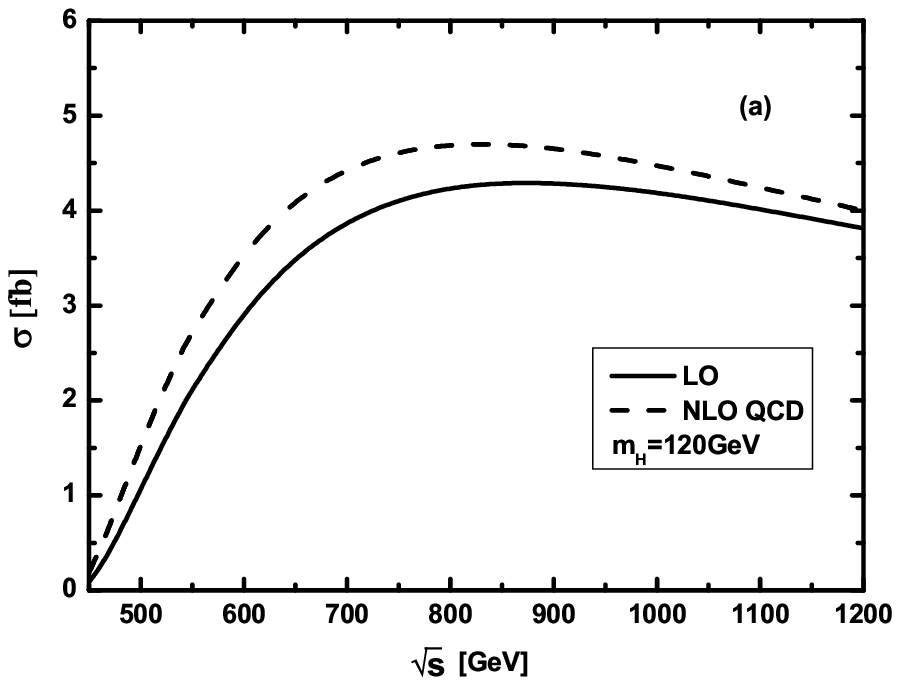}
\includegraphics[scale=0.75]{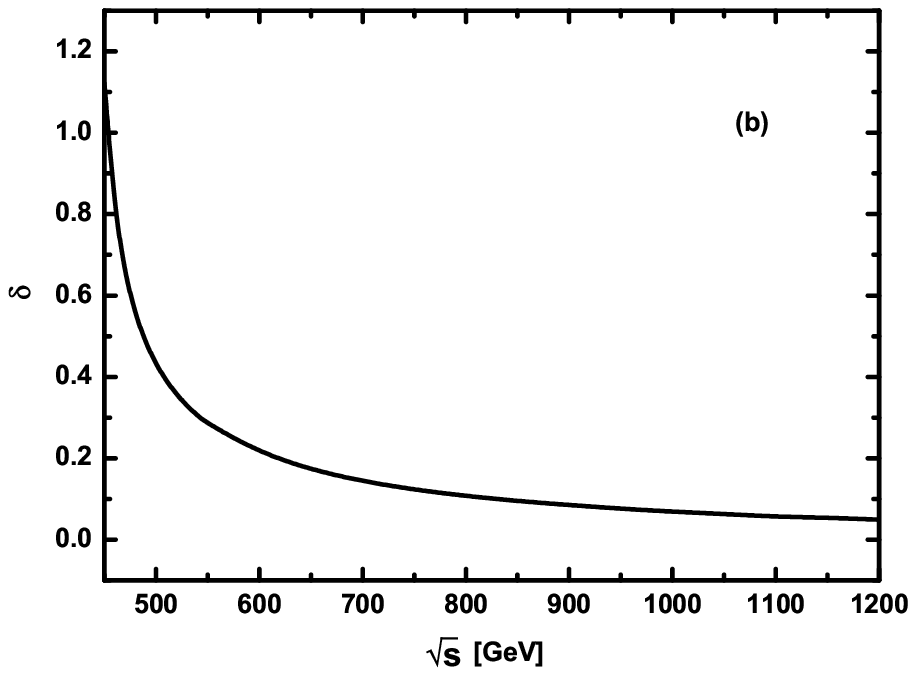}
\caption{\label{fig5} (a) The LO and QCD corrected cross sections
for the process \eettz as the functions of colliding energy
$\sqrt{s}$ with $m_H=120~GeV$. (b) The corresponding relative QCD
radiative correction($\delta^{QCD}$) versus $\sqrt{s}$.}
\end{figure}

In Fig.\ref{fig5}(a) we present the LO and ${\cal O}(\alpha_{s})$
QCD corrected cross sections($\sigma_0$,
$\sigma^{QCD}\equiv\sigma_0+ \Delta\sigma^{QCD}$) as the functions
of colliding energy $\sqrt{s}$ with $m_H=120~GeV$. The
corresponding relative QCD radiative
correction($\delta^{QCD}\equiv \frac{
\Delta\sigma^{QCD}}{\sigma_{0}}$) is presented in
Fig.\ref{fig5}(b). We can see from Figs.\ref{fig5}(a-b) that both
the LO and QCD corrected cross sections are sensitive to the
colliding energy when $\sqrt{s}$ is less than $700~GeV$, while
decrease slowly when $\sqrt{s}>900~GeV$. Fig.\ref{fig5}(b) shows
that the QCD relative radiative correction has a large value in
the vicinity where the colliding energy is close to the $t\bar
tZ^0$ threshold due to Coulomb singularity effect.

\par
\begin{figure}
\centering
\includegraphics[scale=0.75]{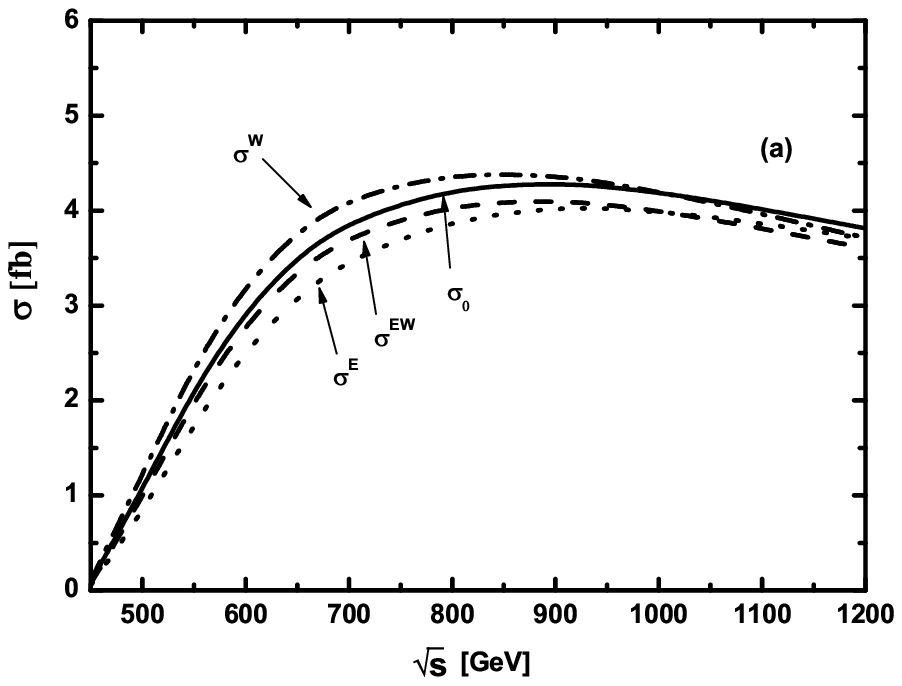}
\includegraphics[scale=0.75]{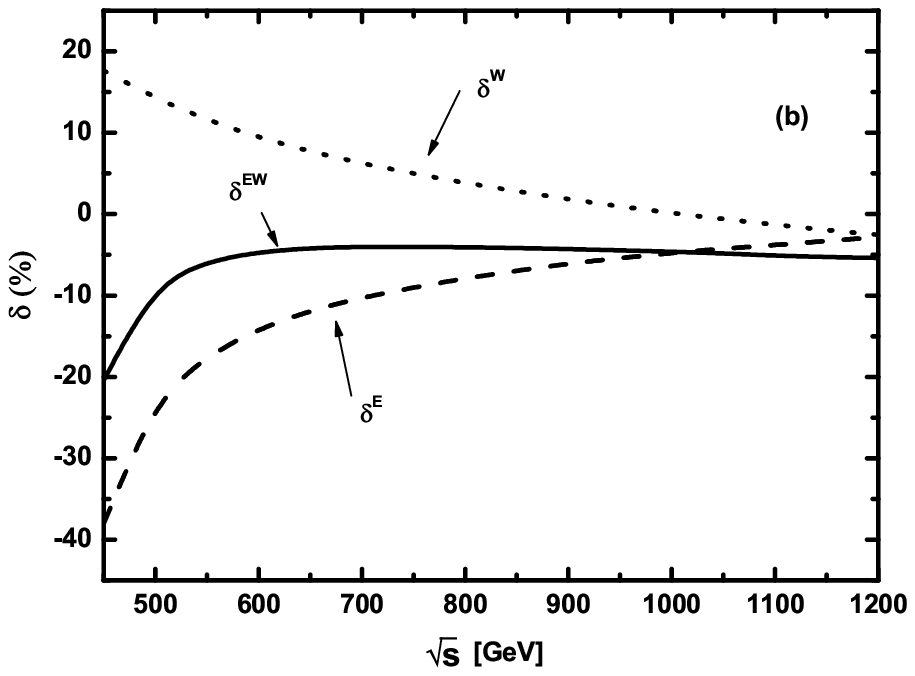}
\caption{\label{fig6} (a) The LO cross section, QED, weak, and total EW
corrected cross sections($\sigma^{E,W,EW}$) for the process
\eettz as the functions of colliding
energy $\sqrt{s}$ with $m_H=120~GeV$. (b) The corresponding
relative QED, weak, and total EW radiative corrections ($\delta^{E,W,EW}$) versus
$\sqrt{s}$.}
\end{figure}

In Figs.\ref{fig6}(a-b) we depict the curves of the LO cross
section($\sigma_0$), ${\cal O}(\alpha_{ew})$ QED, weak, and total EW
corrected cross
sections($\sigma^{E,W,EW}\equiv\sigma_0+\Delta\sigma^{E,W,EW}$) as
the functions of colliding energy $\sqrt{s}$ with $m_H=120~GeV$, and
the corresponding relative radiative
corrections($\delta^{E,W,EW}\equiv \frac{\Delta\sigma^{E,W,EW}}
{\sigma_{0}}$) versus colliding energy are drawn in
Fig.\ref{fig6}(b). Figs.\ref{fig6}(a-b) show that the QED and total
EW corrections always suppress the LO cross section, while the weak
correction part enhances the LO cross section, except it slightly
suppresses the LO cross section in the region of
$\sqrt{s}>1020~GeV$. The QED relative correction is about $-38.0\%$
at $t\bar t Z^0$ threshold and rises to $-2.8\%$ at $1.2~TeV$, but
the weak relative correction is about $17.6\%$ at threshold and
decreases to $-2.6\%$ at $1.2~TeV$. Thus the QED and weak
contributions partially compensate each other when they combine to
form the total EW correction, and yield the EW relative corrections
of about $-20.4\%$ at threshold and $-5.4\%$ at $1.2~TeV$. Again, we
see the Coulomb singularity effects on the curves of corrections
$\delta^E$ and $\delta^{EW}$ in Fig.\ref{fig6}(b) in the vicinity of
threshold. There the curves for the QED, weak, and total EW relative
corrections in Fig.\ref{fig6}(b) demonstrate that the large EW
corrections near the threshold region are mainly contributed by QED
corrections which originate from the diagrams involving an
instantaneous virtual photon in loop with a small spatial momentum.
To show the numerical results presented in Figs.\ref{fig5}(a-b) and
Figs.\ref{fig6}(a-b) more precisely, we list some typical numerical
results for Born cross section($\sigma_{0}$), QCD, EW corrected
cross sections($\sigma^{QCD}$, $\sigma^{EW}$) and QCD, EW relative
corrections($\delta^{QCD,EW}\equiv
\Delta\sigma^{QCD,EW}/\sigma_{0}$) for the process \eettz in Table
\ref{tab2}. All these results show that when $\sqrt{s}$ goes up from
$500~GeV$ to $1.2~TeV$, the relative QCD(EW) radiative correction,
$\delta^{QCD}$($\delta^{EW}$), varies from $43.16\%$($-8.63\%$) to
$4.99\%$($-5.42\%$).
\begin{table}
\begin{center}
\begin{tabular}{|c|c|c|c|c|c|}
\hline $\sqrt{s}$ (GeV)  & $\sigma_{0}(fb)$
& $\sigma^{QCD}(fb)$ & $\delta^{QCD}($\%$)$& $\sigma^{EW}(fb)$ & $\delta^{EW}($\%$)$  \\
\hline 500  & 1.0648(2)  & 1.5244(4)  & 43.16(3) & 0.9729(6)  & -8.63(5) \\
\hline 600  & 2.9432(4)  & 3.5741(7)  & 21.46(2) & 2.805(2)   & -4.70(7) \\
\hline 800  & 4.243(1)   & 4.698(2)   & 10.72(3) & 4.070(4)   & -4.07(7) \\
\hline 1000 & 4.187(1)   & 4.478(2)   &  6.93(4) & 3.994(4)   & -4.63(8) \\
\hline 1200 & 3.811(1)   & 4.001(2)   &  4.99(5) & 3.604(4)   & -5.42(8) \\
\hline
\end{tabular}
\end{center}
\begin{center}
\begin{minipage}{15cm}
\caption{\label{tab2} The numerical results of the tree-level,
one-loop QCD, EW corrected cross sections and the relative QCD, EW radiative
corrections($\delta^{QCD,EW}\equiv \Delta\sigma^{QCD,EW}/\sigma_{0}$) for
the process \eettz, by taking  $m_H=120~GeV$ and $\sqrt{s}=500~GeV$,
$600~GeV$, $800~GeV$, $1000~GeV$, $1200~GeV$, separately. }
\end{minipage}
\end{center}
\end{table}

\par
\begin{figure}
\centering
\includegraphics[scale=0.5]{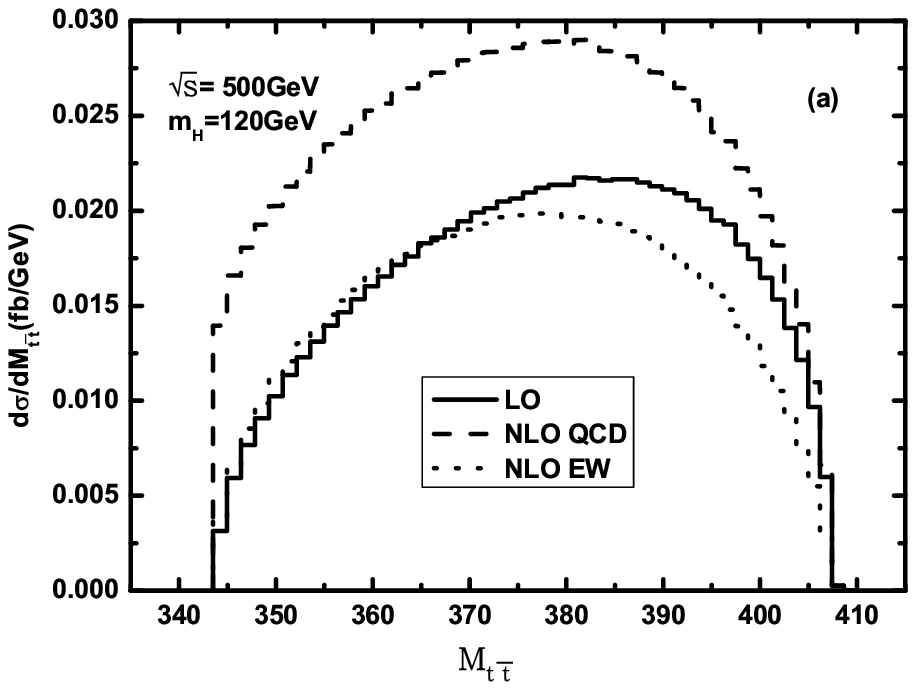}
\includegraphics[scale=0.5]{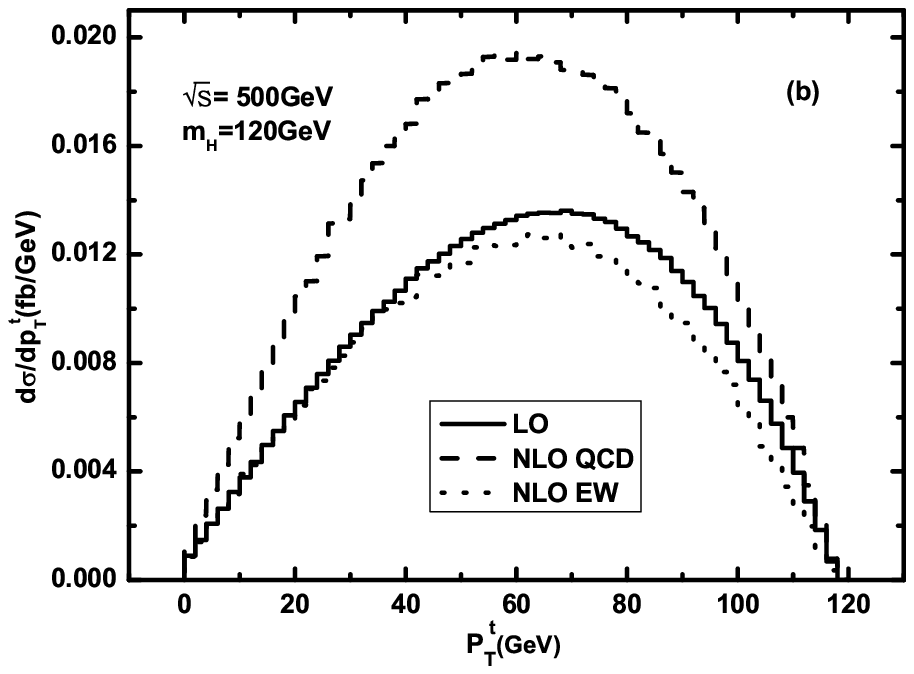}
\includegraphics[scale=0.5]{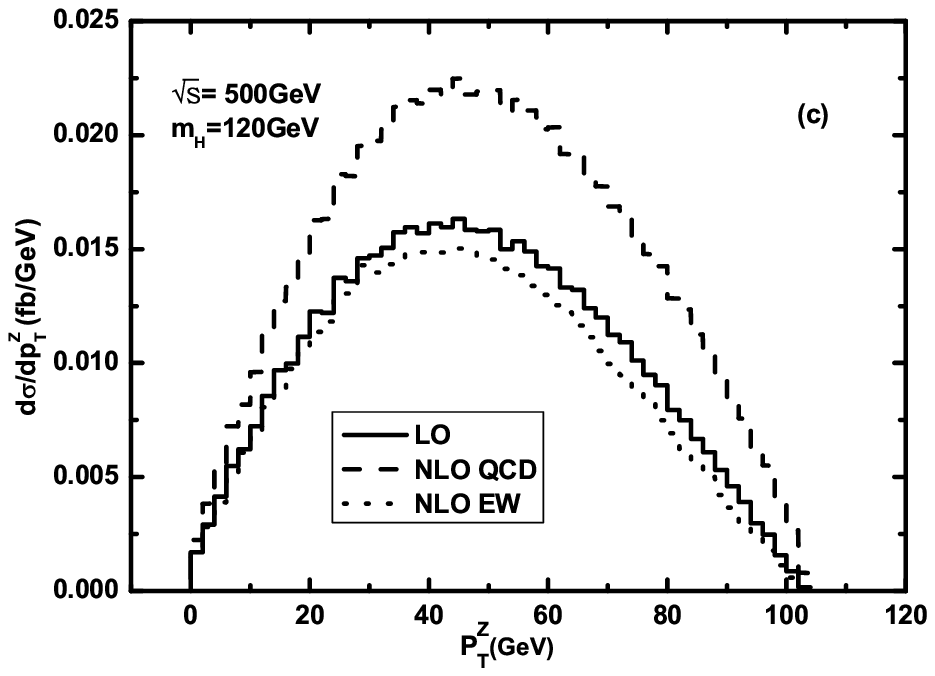}
\caption{\label{fig7} The spectrums for \eettz process at LO, QCD
and EW one-loop order in conditions of $m_H=120~GeV$ and
$\sqrt{s}=500~GeV$. (a) The spectrums of invariant mass of
top-quark pair $M_{t\bar t}$. (b) The spectrums of the transverse
momentum of top-quark $p_T^t$. (c) The  spectrums of the transverse
momentum of $Z^0$-boson $p_T^Z$.  }
\end{figure}

In Fig.\ref{fig7}(a) we present the spectrums of the invariant
mass of top-quark pair($M_{t\bar t}$) at LO, QCD(EW) one-loop
order for the process \eettz by taking $m_H=120~GeV$ and
$\sqrt{s}=500~GeV$. The distributions of the transverse momenta of
top-quark($d\sigma_0/d p_T^t$, $d\sigma^{QCD,EW}/d p_T^t$) and
$Z^0$-boson($d\sigma_0/d p_T^Z$, $d\sigma^{QCD,EW}/d p_T^Z$) are
depicted in Fig.\ref{fig7}(b) and (c), separately.
Fig.\ref{fig7}(a) shows that the QCD corrections always
significantly enhance the LO differential cross section
$d\sigma_0/d M_{t \bar t}$ in the whole plotted region, except the
enhancement becomes much smaller when $M_{t\bar t}>400~GeV$, while
the EW correction either slightly enhances the LO differential
cross section of $d \sigma_0/d M_{t\bar t}$ in the region of
$M_{t\bar t}<365~GeV$, or distinctly suppresses the differential
cross section when $M_{t\bar t}>375~GeV$ as shown in
Fig.\ref{fig7}(a). All these three figures demonstrate that the
QCD corrections induce the enhancement to the LO differential
cross sections of the invariant mass of top-quark pair, the
transverse momenta of top-quark and $Z^0$-boson. We can see from
Figs.\ref{fig7}(b) and (c) that the EW radiative corrections can
only slightly reduce the LO differential cross sections of $d
\sigma_0/d p_T^t$ and $d \sigma_0/d p_T^Z$ in all possible $p_T^t$
and $p_T^Z$ regions.

\vskip 10mm
\section{Summary}
\par
Probing precisely the top physics and searching for the signature of
new physics are important tasks in present and future high energy
physics experiments. The future $e^+e^-$ linear collider could
provide much more efficient laboratory to put these measurements
into practice with a cleanest environment. In this paper we have
studied the full one-loop QCD and EW corrections to the \eettz
process in the SM. We investigate the dependence of the effects
coming from the QCD and EW contributions to the cross section of
process \eettz on colliding energy $\sqrt{s}$. We find that the QCD
correction enhances the tree-level cross section, while the one-loop
EW correction suppresses the LO cross section. Our numerical results
show that when $m_H=120~GeV$ and colliding energy has the values of
$500~GeV$ and $1.2~TeV$, the corresponding relative QCD(EW)
corrections are $43.16\%(-8.63\%)$ and $4.99\%(-5.42\%)$,
respectively. We present also the LO, QCD and EW corrected
differential cross sections of transverse momenta of final top-quark
and $Z^0$-boson, and the distributions of top-quark pair invariant
mass. We conclude that both the QCD and EW radiative corrections
have relevant impact on the \eettz process, and should be included
in any reliable analysis.

\vskip 10mm
\par
\noindent{\large\bf Acknowledgments:} This work was supported in
part by the National Natural Science Foundation of
China(No.10575094), the National Science Fund for Fostering
Talents in Basic Science(No.J0630319), Specialized Research Fund
for the Doctoral Program of Higher
Education(SRFDP)(No.20050358063) and a special fund sponsored by
Chinese Academy of Sciences.

\end{document}